\documentclass[aps,twocolumn,superscriptaddress,prl,floatfix]{revtex4}
\usepackage{amssymb}
\usepackage{amsmath}
\usepackage{mathtools}
\usepackage{graphicx}
\usepackage{xfrac}
\usepackage[dvipsnames]{xcolor}
\usepackage{gensymb}
\usepackage{bm}

\begin{document}
\title{From anyons to Majoranas}

\author
{Jay Sau,$^{1,2,3}$ Steven Simon,$^{4}$ Smitha Vishveshwara,$^{5}$ James R. Williams$^{1,2,6 \ast}$\\
\vspace*{0.25cm}
\small{$^{1}$\emph{Department of Physics, University of Maryland, College Park, MD, USA}}\\
\small{$^{2}$\emph{Joint Quantum Institute, University of Maryland, College Park, MD, USA}}\\
\small{$^{3}$\emph{Condensed Matter Theory Center, University of Maryland, College Park, MD, USA}}\\
\small{$^{4}$\emph{Rudolf Peierls Centre for Theoretical Physics, Oxford University, Oxford, OX1 3PU, UK}}\\
\small{$^{5}$\emph{Department of Physics, University of Illinois at Urbana-Champaign, Urbana, IL, USA}}\\
\small{$^{6}$\emph{Quantum Materials Center, University of Maryland, College Park, MD, USA}}\\
\footnotesize{$^\ast$\emph{To whom correspondence should be addressed; E-mail:  jwilliams@physics.umd.edu.}}\\
}

\date{\today}

\begin{abstract}
Anyons, particles that are neither bosons nor fermions, were predicted in the 1980s, but strong experimental evidence for the existence of the simplest type on anyons has only emerged this year. Further theoretical and experimental advances promise to nail the existence of more exotic types on anyons, such as Majorana fermions, which would make topological quantum computation possible. 
\end{abstract}
\maketitle

Progress in condensed matter physics can be described with an appropriate term borrowed from evolutionary biology: punctuated equilibrium. This term is used to describe abrupt jumps in species evolution that are separated by long periods, known as stasis, having little or no apparent change. In the early 1980s a paradigm shift occurred in condensed matter with the discovery of fractional quantum Hall effect and the theoretical prediction that such systems could, as an emergent phenomenon, harbor anyons -- particles that are neither bosons nor fermions.  Thereafter both experiments and theory developed at a slow pace for a long period. Almost forty years later, these developments have finally culminated in two beautiful experiments that together provide the strongest experimental proof to date that anyons do exist~\cite{Bartolomei20, Nakamura20}.  

Each experiment detects anyons of the simplest variety, in that they acquire a fractional phase that interpolates between bosons and fermions. One experiment measures particle correlations. This technique measures the degree to which particles to like to bunch together: bosons bunch together, fermions like to stay apart and anyons do something in between. The other uses interferometry to pinpoint the phase obtained by encircling a particle around another. This experiment utilizes the exchange properties of particles. Interchanging coordinates of two bosons adds a quantum mechanical phase of 2 to the total wave function, while for two fermions its pi and for two anyons it is somewhere in between. In addition to these simpler anyons, quantum Hall systems hold promise for the realization of more exotic anyons, such as Majorana fermions, which are sensitive to the order in which they are braided -- a property that could enable certain schemes for quantum computation~\cite{Nayak08}.

The Majorana fermion, which is its own antiparticle, was proposed in 1937 and for a long time it seemed irrelevant to condensed matter physics.  At the turn of the 21st century theory predicted~\cite{Kitaev01, Read00} that Majoranas could also occur in condensed matter systems.  In 2012 the first experimental signatures of Majorana fermions came to light in the context of topological superconducting nanowire devices~\cite{Mourik12}, which have similar physics to quantum Hall systems.  Yet these signatures have so far failed to develop into a definitive confirmation. After many frustrating partial results, it may appears that the field has now settled into period of stasis.  However,  there have been slow advances on several other fronts, any one of which could be the key the next big jump. 

We will discuss two such particularly promising, interconnected advances, but to understand their significance, we should first visit a subtle point in the field of superconductivity that places a historic concept in a modern setting: the distinction between the Bose-Einstein condensate (BEC) regime and the conventional Bardeen-Cooper-Schrieffer (BCS) superconductivity regime. The BEC regime is characterized by superconducting pairing that is strong enough that the energy needed to break a pair (the pair-breaking gap $\Delta$) is comparable to the Fermi energy ($E_F$). In this regime, extremely strong attractive interactions can, in principle, bind pairs of electrons into molecules. These molecules, being bosons, can then condense to form a BEC. Although this so-called molecular theory of superconductivity explained many properties of superconductors, it could not be applied to most superconductors, for which the pairing interaction is weak.  Since the development of the BCS theory, a select few systems have been found with pairing interactions strong enough to be close to the BEC regime.  These include SrTiO$_3$ , high-T$_C$ superconductors, ultra-cold atomic gases, and most recently, iron-based superconductors.

One particular feature of interest in this regime -- though only practically accessible in the BEC regime -- are the so-called Caroli-de Gennes-Matricon (CdGM) states in the core of a superconducting vortex. In the BCS regime, the high current density in the core of vortices is sufficient to locally destroy superconducting pairing.  The almost metallic excitations in the vortex core then display particle-in-a-box physics, confined by the surrounding superconductor, with discrete energy levels having energy spacing of $\Delta^2/E_F$  (see Fig. 1a) which is far too small to be measured in conventional BCS superconductors. In contrast, the `molecular' pairing deep in the BEC limit remains robust in the vortex core, which has almost no low-lying excitations in this case. The discrete energy levels in the vortex core reappear in the cross-over between the BEC and the BCS regimes where the energy spacing is comparable to the superconducting gap which in principle is large enough to be measured. Unfortunately, measuring CdGM states is challenging as the few exotic superconductors near the BEC regime all have their own difficulties.  For example, high-TC superconductors, despite having very strong pairing, have gapless (or nodal)  superconductivity, which prevents the formation of proper CdGM states since the particles-in-the-box can leak into the bulk of the superconductor.  Thus, the observations of CdGM states in some iron superconductors~\cite{Chen18, Kong19} have been a triumph.  These superconductors may be the just the right (`Goldilocks') materials which have strong pairing and no other show-stopping problems, providing a flexible family of materials that are fully gapped in the bulk and can be tuned close to the BEC regime.

\begin{figure}[t!]
\center \label{fig1}
\includegraphics[width=2.5 in]{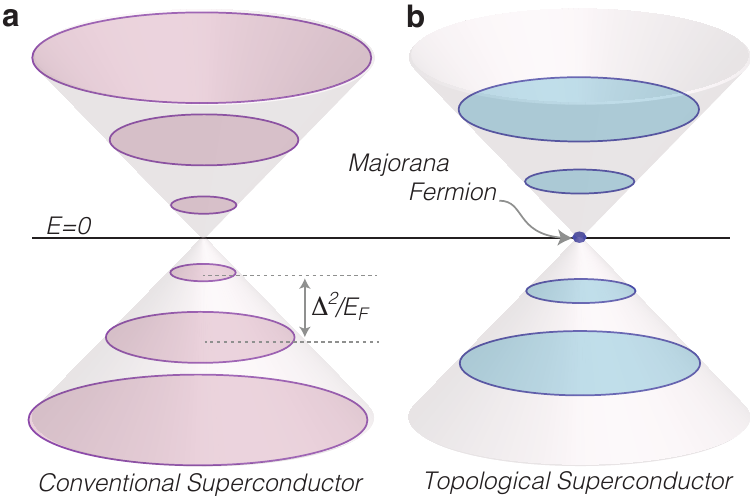}
\caption{\footnotesize{\textbf{Comparison of states localized inside a vortex.} \textbf{a}, In conventional superconductors, states localized inside a vortex are spaced by $\Delta^2/E_F$, where $\Delta$ is a metric of the strength of the superconducting pairing strength and $E_F$ is the Fermi energy. These states are situated at equal, but opposite, energies around $E=0$, a result of the particlehole symmetry provided by the superconductor. The lowest energy state is displaced from $E=0$. \textbf{b}, States inside a vortex in a topological superconductor are also spaced by $\Delta^2/E_F$, but include a state pinned to $E=0$. This state is a Majorana fermion.  }}
\end{figure}

The observation of CdGM states holds particular significance in the context of topological superconductivity. In any superconductor, the set of CdGM states are particle-hole symmetric, meaning that if there is a state for electron addition at energy $E$ there is also a state for electron removal at energy $-E$.  This symmetry is assured by the fact that adding an electron is essentially the same as removing an electron and adding a pair of electrons (a boson) from/to the condensate. What is special about topological superconductors is that one of the CdGM states sits at zero energy (see Fig. 1b). The zero-energy state is unique in that it must be its own partner under particle-hole symmetry, meaning it is a Majorana. Such a Majorana mode cannot move away from zero energy because any continuous deformation of the spectrum that moves the Majorana away from either a positive or negative energy would break the required particle-hole symmetry inherent to superconductivity. It is also this symmetry which underlies the topological protection of Majorana qubits, should they be found in experiment. At least some of the iron-based superconductors are prime suspects for being topological superconductors which gives hope that the observed zero-energy state in a vortex core could be the long-sought Majorana.

The conclusive identification of excitations in vortices may provide the key technological advance needed for topological quantum computation~\cite{November19}. Vortices on the surface of a superconductor can be manipulated: braiding two vortices and then fusing them together produces quantum states that depend on how the braiding was performed.  This process is essential to determine the anyonic properties of a Majorana and in the creation of a topological qubit. This type of manipulation could potentially be assimilated into a scalable device architecture because in addition to bulk superconductors, vortices can also exist in Josephson junctions.  These junctions are superconducting devices where two superconductors are coupled via a weak link.  Unlike bulk vortices, around which supercurrents flow in full circle, these Josephson vortices are located at points along the extended junction where the current switches direction. These points can be controlled with external knobs such as current pulses and magnetic fields. Multiple candidate topological materials can be used as the weak link in Josephson junctions, offering prospects for creating entire networks in which vortices can be manipulated via local fields and current pulses~\cite{Hegde19}. In this way Majorana-based devices could be used for quantum information processing. As with its original nanowire cousins, the Josephson junction system is a scalable platform, allowing for the coupling of many Majorana-based qubits needed for topological quantum computation.


\begin{thebibliography}{10}
\bibitem{Bartolomei20} H. Bartolomei et al, ``Fractional statistics in anyon collisions'', Science 368, 173-177 (2020). 
\bibitem{Nakamura20} J. Nakamura, S. Liang, G. C. Gardner, M. Manfra. ``Direct Observation of anyonic braiding statistics'', Nature Phys. 16, 931-936 (2020).  
\bibitem{Nayak08} C. Nayak et al., ``Non-Abelian anyons and topological quantum computation'', Rev. Mod. Phys. 80, 1083 (2008).
\bibitem{Kitaev01} A. Kitaev. ``Unpaired Majorana Fermions in Quantum Wires". Phys.-Usp.44, 131-136 (2001).\bibitem{Read00} N. Read and D. Green. ``Paired states of fermions in two dimensions with breaking of parity and time-reversal symmetries and the fractional quantum Hall effect". Phys. Rev. B.61, 10267 (2000).
\bibitem{Mourik12} V. Mourik et al. ``Signatures of Majorana Fermions in Hybrid Superconductor-Semiconductor Nanowire Devices", Science 336, 1003-1007 (2012).
\bibitem{Chen18} M. Chen et al. ``Discrete energy levels of Caroli-de Gennes-Matricon states in quantum limit in FeTe$_{0.55}$Se$_{0.45}$'', Nature Comm. 9, 970 (2018).
\bibitem{Kong19} L. Kong, et al. ``Half-integer level shift of vortex bound states in an iron-based superconductor'', Nature Phys. 15, 1181-1187 (2019). 
\bibitem{November19} B. H. November, J. D. Sau, J. R. Williams, J. E. Hoffman. ``Scheme for Majorana Manipulation Using Magnetic Force Microscopy'', arXiv:1905.09792 (2019).
\bibitem{Hegde19} S. S. Hegde et al. ``A topological Josephson junction platform for creating, manipulating and braiding Majorana bound states'', Annals of Physics 423, 168326 (2020). 
\end{thebibliography}
\end{document}